\documentclass[journal,10pt]{IEEEtran}
\usepackage{etoolbox}
\usepackage{dtsc-creafig}
\usepackage{notation}

\usepackage{epsfig}
\usepackage{color}
\usepackage{amsmath}
\usepackage{amssymb}
\usepackage{mathabx}
\usepackage{enumerate}
\usepackage{multirow}
\usepackage{bbm}
\usepackage{algorithm}
\usepackage{algorithmic}
\usepackage{lipsum,amsmath,multicol}
\usepackage{stfloats}
\usepackage{arydshln} 
\usepackage{amsfonts}
\usepackage{dsfont}
\usepackage{upgreek}
\usepackage{pgfplots} 
\usepackage{caption}
\usepackage{subfigure} 
\usepackage{booktabs} 

\newcommand{\newt}[1]{\textcolor{black}{#1}}
\newcommand{\noteISV}[1]{\textcolor{black}{#1}}

\usepackage{cite}

\usepackage{acronym}
\acrodef{LMMSE}{linear minimum mean square error}
\acrodef{FIR}{finite impulse response}
\acrodef{MIMO}{multiple-input multiple-output}
\acrodef{IID}{independent and identically distributed} 
\acrodef{S/P}{serial to parallel} 
\acrodef{BP}{belief propagation}
\acrodef{SPA}{sum-product algorithm}     
\acrodef{GMP}{Gaussian message passing} 
\acrodef{KL}{Kullback-Leibler} 
\acrodef{ML}{maximum likelihood}
\acrodef{pdf}{probability density function}
\acrodef{SISO}{single-input single-output}
\acrodef{DFE}{decision feedback equalization}
\acrodef{EXIT}{extrinsic information transfer}
\acrodef{BER}{bit error rate}
\acrodef{SER}{symbol error rate}
\acrodef{ECC}{error correction code}
\acrodef{FEC}{forward error correction}
\acrodef{MSE}{mean-square-error}
\acrodef{MMSE}{minimum mean square error}
\acrodef{TLMMSE}{turbo linear minimum mean square error}
\acrodef{APP}{a posteriori probabilities}
\acrodef{LLRs}{log-likelihood ratios}
\acrodef{LLR}{log-likelihood ratio}
\acrodef{BEP}{block expectation propagation}
\acrodef{nuBEP}{non-uniform block expectation propagation}
\acrodef{TBEP}{turbo block expectation propagation}
\acrodef{ISI}{intersymbol interference}
\acrodef{ICI}{intercarrier interference}
\acrodef{AWGN}{additive white Gaussian noise}
\acrodef{MAP}{maximum a posteriori}
\acrodef{LDPC}{low-density parity-check}
\acrodef{EP}{expectation propagation}
\acrodef{SD}{sphere decoding}
\acrodef{MCMC}{Markov chain Monte Carlo}
\acrodef{GTA}{Gaussian tree approximation}
\acrodef{CHEMP}{channel hardening-exploiting message passing}
\acrodef{BP}{belief propagation}
\acrodef{LTI}{linear time invariant}
\acrodef{pmf}{probability mass function}
\acrodef{CSI}{channel state information}

\ifCLASSINFOpdf
\else
\fi
\begin{document}
%
\title{{Self and turbo iterations for MIMO receivers and large-scale systems}}
%
%
%

\author{Irene~Santos and 
        Juan~Jos\'e~Murillo-Fuentes
\thanks{I. Santos and J.J. Murillo-Fuentes are with the Dept. Teor\'ia de la Se\~nal y Comunicaciones, Universidad de Sevilla, Camino de los Descubrimientos s/n, 41092 Sevilla, Spain. E-mail: {\tt \{irenesantos,murillo\}@us.es}}
\thanks{This work was partially funded by the Spanish government (TEC2016-78434-C3-R) and the European Union (MINECO/FEDER, UE).}}

\maketitle

\begin{abstract}
We investigate a turbo soft detector based on the \ac{EP} algorithm for large-scale \ac{MIMO} systems. 
Optimal detection in \ac{MIMO} systems becomes computationally unfeasible for high-order modulations and/or large number of antennas. In this situation, the \ac{LMMSE}  exhibits a low-complexity with a good performance, however far from optimal. To improve the performance, the \ac{EP} algorithm can be used. In this paper, we review previous EP-based detectors and \notajj{enhance} their estimation in terms of complexity and performance. Specifically, \notaI{we improve the convergence of the self-iterated EP stage by replacing the uniform prior by a non-uniform one, which better characterizes the information returned by the decoder once the turbo procedure starts. }
We also review the EP parameters to avoid instabilities when using high-order modulations and to reduce the \notajj{computational} complexity. Simulation results illustrate the robustness and \notajj{enhanced} performance of this \notaI{novel} detector in comparison with previous approaches found in the literature.  \noteISV{Results also show that the proposed detector is robust in the presence of imperfect \ac{CSI}. }
\end{abstract}

\begin{IEEEkeywords}
Expectation propagation (EP), MMSE, low-complexity, MIMO, turbo detection, feedback.
\end{IEEEkeywords}

%
\IEEEpeerreviewmaketitle

\section{Introduction}
%
%
%
%


\notajj{Multiple-input multiple-output (\ac{MIMO}) antennas are of great interest in communications systems due, among others, }to the need of transmitting at high rates \cite{Rusek13}. 
\notajj{In \ac{MIMO} detection we aim at estimating the symbols transmitted by the transmit antennas using the output of the receive ones. }
This estimation can be probabilistic, resulting in a high benefit for modern channel decoders. In addition, the performance can be \notajj{further} improved with a turbo detection scheme, i.e., by exchanging information between the decoder and the soft detector iteratively. 

Optimal detectors, such as the \ac{MAP} algorithm, suffer from an intractable computational complexity for high order constellations and/or large number of \notajj{transmit} antennas. In this \notajj{scenario, non-optimal  approximate solutions are used instead}. 
%
The \ac{SD} method provides an approximated marginal posterior \ac{pdf} in a subspace of the whole set of possible transmitted words given by the constellation \cite{Studer08}. Another alternative to approximate the posterior distribution is the use of \ac{MCMC} algorithms \cite{Datta13}. However, their required complexity to obtain an accurate enough performance remains unfeasible for large scale scenarios. 


\notajj{The linear minimum mean square error (\ac{LMMSE}) is a quite extended solution due to its low computational} complexity. 
Since its performance is far from the optimal, alternative approaches can be found in the literature. The \ac{GTA} algorithm \cite{Goldberger11} \noteISV{firstly ignores the discrete nature of symbols to} construct a tree-factorized Gaussian approximation to the \ac{APP}. Then, \noteISV{it} estimate\noteISV{s} the marginals with \ac{BP}. The \ac{CHEMP}  \cite{Narasimhan14} is a message-passing algorithm where all the exchanged messages are approximated by Gaussian distributions. \noteISV{However, its performance degrades when using high-order modulations \cite{Cespedes17,Cespedes17thesis}. }

\notajj{Recently, the expectation propagation (\ac{EP}) algorithm \cite{Minka01thesis,Seeger05} has been proposed as better approach to approximate the posterior. }
%
This algorithm has been already applied to equalization \cite{Santos16,Santos17,Santos18} and  \ac{MIMO} detection \cite{Cespedes14,Cespedes17,Cespedes17thesis}. 
In these works, it was showed that the \ac{EP} detector improved the performance of \ac{LMMSE}, \ac{GTA} and \ac{CHEMP} with complexity proportional to the \ac{LMMSE} algorithm. \notaI{The extension to MIMO turbo detection is introduced in \cite{Cespedes17thesis,Senst11}. Approach \cite{Cespedes17thesis} is based on a self-iterated EP approach implemented with a damping procedure, although uniform priors are employed to describe the information from the decoder. On the other hand, \notajj{the method in} \cite{Senst11} does not include self-iterations within the EP stage, but assumes non-uniform priors instead. Since a non-uniform prior for the symbols better characterizes the information returned by the decoder, approach \cite{Senst11} improves the performance in \cite{Cespedes17thesis}. }

%

In this paper, we focus on the interaction between self and turbo iterations, \notaI{outperforming both \cite{Cespedes17thesis,Senst11} approaches. }
\notaIb{We use non-uniform priors distributed according to the channel decoder output during the moment matching procedure, borrowing from \cite{Senst11}. Then, this procedure is repeated \notajj{at the self-iterated EP stage} \noteISV{to update} its estimation, including a damping procedure \notajj{similar to the one proposed in} \cite{Cespedes17thesis}. }Following the guidelines in \cite{Santos18}, we also optimize the EP parameters to avoid instabilities with high-order modulations and reduce the computational complexity of the algorithm. This novel solution outperforms \cite{Senst11} due to the self-iterated EP approach and the damping procedure. For large constellations this self-iterated EP approach greatly improves convergence compared to the LMMSE.




\section{System Model}\LABSEC{sysmod}


The turbo architecture of a soft \ac{MIMO} detector where $\Ntx$ transmit
antennas communicate to a receiver with $\Nrx$ antennas can be divided into three parts. 


\subsection{Transmitter}


The information bit vector, $\vect{a}=[\allvect{a}{1}{\k}]\trs$, is encoded  into the codeword $\vect{b}=[\allvect{b}{1}{\t}]\trs$ with a code rate $\rate=\k/\t$. The codeword is partitioned into $\tamframe=\lceil \t/ \log_{2}\modsize \rceil$ blocks, $\vect{b}=[\vect{b}_{1}, ..., \vect{b}_{\tamframe}]\trs$, where  $\vect{b}_{\iter}=[b_{\iter,1}, ..., c_{\iter,Q}]$ and $Q=\log_2(\modsize)$. 
\newt{Every $\vect{b}_{\iter}$ is modulated to get one out of $\modsize$ possible complex-valued symbols \noteISV{that belong to the alphabet $\mathcal{A}$}.}
The \newt{modulated} symbols are partitioned into $\blocks$ blocks of length $\Ntx$, \newt{$[\vect{\beforechannel}[1],...,\vect{\beforechannel}[\blocks]]\trs$}, where $\vect{\beforechannel}[\itermimo]=[\beforechannel_{\itermimo,1},...,\beforechannel_{\itermimo,\Ntx}]$. 
Each block is demultiplexed into $\Ntx$ substreams
 through the \ac{S/P} converter. 
Then, the block frames are transmitted over the channel. \newt{Hereafter, we focus on the transmission and estimation of any block $\vect{\beforechannel}[\itermimo]$ and, to keep the notation uncluttered, we will omit the index $\itermimo$.
}

\subsection{Channel model}
The channel is completely specified by \notajj{the known noise variance, $\sigma_\noise^2$, and the weights between each transmitting and receiving antenna, $h_{k,j}$, where $k=1,...,\Nrx$ and $j=1,...,\Ntx$, \notaI{with $\Nrx\geq\Ntx$}}. 
The received signal \newt{ for a channel use, $\vect{\afterchannel}=[\afterchannel_1,...,\afterchannel_{\Nrx}]\trs$ }, 
 is given by
\begin{equation}\LABEQ{system}
\newt{\vect{\afterchannel}=\matr{H}\vect{\beforechannel}+\vect{\noise}}
\end{equation}
\notaIb{where $\matr{H}$ is a $\Nrx\times\Ntx$ full matrix where each $h_{k,j}$ element represents the channel weight of $k$th receiving antenna and $j$th transmitting antenna} and \newt{$\vect{\noise}\sim\cgauss{\vect{\noise}}{\matr{0}}{\sigma_\noise^2\matr{I}}$} is a complex-valued \ac{AWGN} vector. 
\notajj{We assume the coherence time to be larger than the block duration. }
\subsection{Turbo receiver}
The posterior probability of the transmitted symbol vector \newt{$\vect{\beforechannel}$} given the whole vector of observations \newt{$\vect{\afterchannel}$} yields\newt{
\begin{align}\LABEQ{pugiveny}
\hspace{-0.2cm}p(\vect{\beforechannel}|\vect{\afterchannel})=\frac{p(\vect{\afterchannel}|\vect{\beforechannel})p_{\noteISV{D}}(\vect{\beforechannel})}{p(\vect{\afterchannel})} 
\propto \; \cgauss{\vect{\afterchannel}}{\matr{H}\vect{\beforechannel}}{\std{\noise}^{2}\matr{I}} \prod_{\iter=1}^{\Ntx} p_D(\beforechannel_{\iter}), 
\end{align}
}
where the true prior returned by the decoder, \newt{$p_D(\beforechannel_{\iter})$}, is clearly non-Gaussian but a non-uniform discrete distribution. If no information is available from the decoder, then the true prior is assumed to be equiprobable. 


The extrinsic distribution computed by the detector is demapped and given to the decoder as extrinsic log-likelihood ratios\footnote{\newt{For $n=(p-1)\Ntx \noteISV{Q}+1,...,(p-1)\Ntx \noteISV{Q}+\Ntx \noteISV{Q}$.}}, \newt{$L_E(b_n)$}. 
The channel decoder computes an estimation of the information bit vector, $\vect{\hat{a}}$, and an extrinsic \ac{LLR} \notajj{of} the coded bits, \newt{$L_D(b_n)$}. 
These LLRs are again mapped and given to the detector as updated priors, $p_D(\vect{\beforechannel})$. This process is repeated iteratively for a given maximum number of iterations, $\nturbo$, or until convergence. 


\section{The block-EP detector}\LABSEC{EP}

The \ac{EP} algorithm provides a feasible approximation to the posterior distribution 
{in \EQ{pugiveny}}, $p(\vect{\beforechannel}|\vect{\afterchannel})$, \newt{by an iteratively estimated Gaussian approximation, $q^{[\ell]}(\vect{\beforechannel})$, where $\ell$ denotes the iteration number. In this approximation, the product of non-Gaussian terms, $p_D(\beforechannel_{\iter})$, in \EQ{pugiveny} is replaced by a product of to be estimated Gaussians, denoted as $t_\iter^{[\ell]}(\beforechannel_\iter)=\cgauss{\beforechannel_\iter}{\mu_{t_\iter}^{[\ell]}}{\sigma_{t_\iter}^{2[\ell]}}$. We next develop the expression for this approximated posterior as it will be needed later on. The approximated posterior factorizes as 
 \vspace{-0.1cm}
\begin{align}\LABEQ{wholeposterior1}
q^{[\ell]}(\vect{\beforechannel})&\; \propto\; p(\vect{\afterchannel}|\vect{\beforechannel}) \prod_{\iter=1}^{\Ntx} t_\iter^{[\ell]}(\beforechannel_\iter)= \\
&=\cgauss{\vect{\afterchannel}}{\matr{H}\vect{\beforechannel}}{\std{\noise}^{2}\matr{I}} \prod_{\iter=1}^{\Ntx} \cgauss{\beforechannel_\iter}{\mu_{t_\iter}^{[\ell]}}{\sigma_{t_\iter}^{2[\ell]}}
\end{align}
and it is distributed according to a Gaussian given by
\begin{align}\LABEQ{wholeposterior}
&q^{[\ell]}(\vect{\beforechannel})\; = \;\cgauss{\vect{\beforechannel}}{{\boldsymbol{\upmu}_{q}^{[\ell]}}}{{\boldsymbol{\Sigma}_{q}^{[\ell]}}} \nonumber\\
&\propto \, \cgauss{\vect{\beforechannel}} {\left(\matr{H}\her\matr{H} \right)\inv\noteISV{\matr{H}\her}\vect{\afterchannel}}  {\std{\noise}^2\left(\matr{H}\her\matr{H} \right)\inv} \cgauss{\vect{\beforechannel}}{\boldsymbol{\upmu}_{t}^{[\ell]}}{{\boldsymbol{\Sigma}_{t}^{[\ell]}}}
\end{align}
%
where {${\boldsymbol{\Sigma}_{t}^{[\ell]}}=\mathrm{diag}([\sigma_{t_1}^{2[\ell]}, \hdots, \sigma_{t_{\Ntx}}^{2[\ell]}])$}, $\boldsymbol{\upmu}_{t}^{[\ell]}=[\mu_{t_1}^{[\ell]}, \hdots, \mu_{t_{\Ntx}}^{[\ell]}]\trs$. 
The mean and covariance of $q^{[\ell]}(\vect{\beforechannel})$ can be computed as (see (A.7) in \cite{Rasmussen06}),
\begin{align}\LABEQ{covEP}
{\boldsymbol{\Sigma}_{q}^{[\ell]}}&=\left(\std{\noise}^{-2}\matr{H\her}\matr{H}+{\left({\boldsymbol{\Sigma}_{t}^{[\ell]}}\right)\inv}\right)\inv, \\
\boldsymbol{\upmu}_{q}^{[\ell]}&={\boldsymbol{\Sigma}_{q}}^{[\ell]}\left(\std{\noise}^{-2}\matr{H\her}\vect{\afterchannel}+{\left({\boldsymbol{\Sigma}_{t}^{[\ell]}}\right)\inv}\boldsymbol{\upmu}_{t}^{[\ell]}\right). \LABEQ{meanEP}
\end{align}
The $\iter$th marginal of $q^{\noteISV{[\ell]}}(\noteISV{\vect{\beforechannel}})$ can be easily computed from this expression as 
$q^{[\ell]}(\beforechannel_\iter)\sim\cgauss{\beforechannel_\iter}{\mu_\iter^{[\ell]}}{\sigma_\iter^{2[\ell]}}$, where \noteISV{$\mu_\iter^{[\ell]}$ is the $\iter$th entry of $\boldsymbol{\upmu}_{q}^{[\ell]}$ and } $\sigma_\iter^{2[\ell]}$ is the $\iter$th diagonal entry of  matrix ${\boldsymbol{\Sigma}_{q}^{\noteISV{[\ell]}}}$.
Bearing these expressions in mind we next face the update of the factors $t_\iter^{\noteISV{[\ell]}}(\beforechannel_k)$ \noteISV{by means of the \ac{EP} algorithm.} 
}

\newt{The \ac{EP} is based on the minimization of the \ac{KL} divergence between the true distribution in \EQ{pugiveny} and its Gaussian approximation in \EQ{wholeposterior}, which corresponds to matching the expected sufficient statistics between both distributions \cite{Bishop06}. Since \EQ{wholeposterior} is Gaussian distributed this is equivalent to matching their means and variances, which is commonly referred to as moment matching. 
Hence,
along $\ell=1,...,\iterep$ iterations, we estimate the new values of its moments, $\mu_{t_\iter}^{[\ell+1]}$ and $\sigma_{t_\iter}^{2[\ell+1]}$, by 
\begin{equation}\LABEQ{MM}
\frac{q^{\noteISV{[\ell]}}(\beforechannel_k)}{t^{\noteISV{[\ell]}}_\iter(\beforechannel_k)}p_D(\beforechannel_\iter) \stackrel{\mbox{\begin{tabular}{c}moment\\matching\end{tabular}}}{\longleftrightarrow} \frac{q^{\noteISV{[\ell]}}(\beforechannel_\iter)}{t^{\noteISV{[\ell]}}_\iter(\beforechannel_k)}t_\iter^{[\ell+1]}(\beforechannel_\iter).
\end{equation}
%
Note that in both terms in \EQ{MM} we have the marginal of the full approximation, $q^{\noteISV{[\ell]}}(\beforechannel_k)$ in \EQ{wholeposterior1}-\EQ{meanEP}, divided by the estimation of the $\iter$th factor, $t^{\noteISV{[\ell]}}_\iter(u_k)$, and then multiplied by the true prior and the new factor, respectively. We define $q_E^{[\ell]}(\beforechannel_\iter)$, that plays the role of an extrinsic marginal distribution, as
\begin{align}\LABEQ{extrinsicnubep}
q_E^{[\ell]}(\beforechannel_\iter)= {q^{[\ell]}(\beforechannel_\iter)}/{t_\iter^{[\ell]}({\beforechannel_\iter})}=
\cgauss{\beforechannel_\iter}{{\mu_{E_\iter}^{{[\ell]}}}}{{\sigma_{E_\iter}^{{2[\ell]}}}}.
\end{align} 
Using \EQ{wholeposterior}-\EQ{meanEP} to compute ${q^{[\ell]}(\beforechannel_\iter)}$ and by the definition of the approximating factors, $t^{\noteISV{[\ell]}}_\iter(\beforechannel_\iter)$, it follows that \newt{(see (A.7) in \cite{Rasmussen06})},
%
\begin{align}\LABEQ{mean_ext}
{\mu_{E_\iter}^{[\ell]}}=\frac{\mu_\iter^{[\ell]}\sigma_{t_\iter}^{2[\ell]}-\mu_{t_\iter}^{[\ell]}\newt{\sigma_\iter^{2[\ell]}}}{\sigma_{t_\iter}^{2[\ell]}-\sigma_\iter^{2[\ell]}}, \;\;\;\;{\sigma_{E_\iter}^{2[\ell]}}=\frac{\sigma_\iter^{2[\ell]}\sigma_{t_\iter}^{2[\ell]}}{\sigma_{t_\iter}^{2[\ell]}-\sigma_\iter^{2[\ell]}} .
\end{align}
Finally, to derive the new moments of $t_\iter^{[\ell+1]}(\beforechannel_\iter)$ from \EQ{MM}, we need to compute \noteISV{the moments of }
\begin{equation}\LABEQ{qEpD}
{p}^{[\ell]}(\beforechannel_\iter)=
q_E^{[\ell]}(\beforechannel_\iter)p_D(\beforechannel_\iter). 
\end{equation}
\noteISV{We will denote its first and second moments as $\mu_{p_\iter}^{[\ell]}$ and $\sigma_{p_\iter}^{2[\ell]}$, respectively. }
We update \noteISV{the factor $t_\iter^{[\ell+1]}(\beforechannel_\iter)$} 
%
with these new values at every iteration $\ell$, using a damping approach, as described in \ALG{MMD}, where the selection of parameters $\epsilon$ and $ \beta$ will be discussed later in this section.
}
The control of negative variances proposed in \cite{Cespedes17,Cespedes17thesis} is also included. \newt{In \TAB{functions} is included a brief description of every function used, with the notation employed for its moments.}

\begin{table}[htb]
\begin{center}
\begin{tabular}{c c l }
\toprule
 &  Mean, Covariance & Description   \\
\midrule
$q^{[\ell]}(\vect{\beforechannel})$ & $\boldsymbol{\upmu}^{[\ell]}_q$, ${\boldsymbol{\Sigma}^{[\ell]}_{q}}$ & Full approximation to the posterior  \\
$q^{[\ell]}({\beforechannel}_\iter)$ & ${\mu}^{[\ell]}_\iter$, ${{\sigma}^{2[\ell]}_{\iter}}$ & Marginal of $q^{[\ell]}(\vect{\beforechannel})$   \\
$t_\iter^{[\ell]}({\beforechannel}_\iter)$ & ${\mu}^{[\ell]}_{t_\iter}$, ${{\sigma}^{2[\ell]}_{t_\iter}}$ & Factors in the approximation $q^{[\ell]}(\vect{\beforechannel})$   \\
$p_D({\beforechannel}_\iter)$ &   & Prior of the transmitted symbols  \\
$q^{[\ell]}_E({\beforechannel}_\iter)$ & ${\mu}^{[\ell]}_{E_\iter}$, ${{\sigma}^{2[\ell]}_{E_\iter}}$ & Extrinsic marginal distribution \\
$p^{[\ell]}({\beforechannel}_\iter)$ & ${\mu}^{[\ell]}_{p_\iter}$, ${{\sigma}^{2[\ell]}_{p_\iter}}$ & Product $q^{[\ell]}_E({\beforechannel}_\iter)p_D({\beforechannel}_\iter)$ \\
\bottomrule
\end{tabular}
\captionof{table}{\small Description of functions used and their parameters, at iteration $\ell$ of \ALG{MMD}. }\LABTAB{functions}
\end{center}
\end{table}

\newt{\subsection{Turbo Detection}}
The whole \ac{EP} procedure for a turbo detector is detailed in \ALG{nuEPalgorithmMIMO}, where \newt{$\ell=1,...,\iterep$ is the iteration number of \ac{EP} and $t=\noteISV{0},...,\nturbo$ is the iteration number of the turbo detection}. \notaI{Unlike \cite{Cespedes17,Cespedes17thesis}, \newt{at Step 2 of this algorithm the priors used in the moment matching (see \EQ{MM}) are the non-uniform \ac{pmf} at the output of the decoder, \newt{$p_D^{[t]}(\beforechannel_\iter)$}}. For this reason, we denote this approach as \ac{nuBEP} detector. }\newt{ Note that for $\nturbo=\noteISV{0}$ we have a standalone version of the detection, with no turbo detection. Also, we may have different values of $\beta$ for each turbo iteration.}


\begin{algorithm}[htb]\newt{
\begin{algorithmic}
\STATE 
{\textbf{Given inputs}}: $p_D(\beforechannel_\iter)$,  $\mu_{t_\iter}^{[\ell]},\sigma_{t_\iter}^{2[\ell]}$, for \noteISV{$\iter=1,\hdots, \Ntx$} and $\vect{\afterchannel}$, $\epsilon$, $\beta$ 
\STATE
1) Compute $q^{\noteISV{[\ell]}}(\vect{\beforechannel})$ in \EQ{wholeposterior}-\EQ{meanEP} and its marginals, $q^{\noteISV{[\ell]}}({u_\iter})$.
\STATE
2) Compute the extrinsic marginal distributions, $q^{\noteISV{[\ell]}}_E(\beforechannel_\iter)$ in \EQ{extrinsicnubep}-\EQ{mean_ext}.
\FOR {$\iter=1,...,\Ntx$}
\STATE
3)  Compute the moments of ${p}^{[\ell]}(\beforechannel_\iter)$ in \EQ{qEpD},  \noteISV{i.e., } 
 $\mu_{p_\iter}^{[\ell]}$ and $\sigma_{p_\iter}^{2[\ell]}$. Set a minimum allowed variance as \mbox{$\sigma_{p_\iter}^{2[\ell]}=\max(\epsilon,\sigma_{{p}_\iter}^{2[\ell]})$}.
 \STATE
 4) Match moments: compute new values of the moments of $t_\iter^{[\ell+1]}(\beforechannel_\iter)$ using \EQ{MM},
\begin{align}
\sigma_{q_{\iter,new}}^{2[\ell+1]}&=\frac{\sigma_{p_\iter}^{2[\ell]}{\sigma_{E_\iter}^{2[\ell]}}}{{\sigma_{E_\iter}^{2[\ell]}}-\sigma_{p_\iter}^{2[\ell]}} , \\
\mu_{q_{\iter,new}}^{[\ell+1]}&= \sigma_{q_{\iter,new}}^{2[\ell+1]}{\left( \frac{\mu_{p_\iter}^{[\ell]}}{\sigma_{p_\iter}^{2[\ell]}}-\frac{{\mu_{E_\iter}^{[\ell]}}}{{\sigma_{E_\iter}^{2[\ell]}}} \right)} \LABEQ{MMvar}. 
\end{align}
\STATE
5) Run damping: Update the values as
\begin{align}
\sigma_{t_\iter}^{2[\ell+1]}&=\left(\beta\frac{1}{\sigma_{q_{\iter,new}}^{2[\ell+1]}} + (1-\beta)\frac{1}{\sigma_{t_\iter}^{2[\ell]}}\right)\inv, \\
\mu_{t_\iter}^{[\ell+1]}&=\sigma_{t_\iter}^{2[\ell+1]}\left(\beta \frac{\mu_{q_{\iter,new}}^{[\ell+1]}}{\sigma_{q_{\iter,new}}^{2[\ell+1]}} + (1-\beta)\frac{\mu_{t_\iter}^{[\ell]}}{\sigma_{t_\iter}^{2[\ell]}}\right). \LABEQ{Lambdak2}
\end{align}
\IF{$\sigma_{\noteISV{t}_{\iter}}^{2[\ell+1]}<0$}
\vspace{-0.3cm}
\STATE
\begin{align}\LABEQ{negativevariance}
\sigma_{t_\iter}^{2[\ell+1]}=\sigma_{t_\iter}^{2[\ell]}, \,\,\,\,\,\,\,\, \mu_{t_\iter}^{[\ell+1]}=\mu_{t_\iter}^{[\ell]}. 
\end{align}
\ENDIF
\STATE
{\textbf{Output}}: $\sigma_{t_\iter}^{2[\ell+1]}, \mu_{t_\iter}^{[\ell+1]}$
\ENDFOR 
\end{algorithmic}}
\caption{Moment Matching and Damping (MMD)}\LABALG{MMD}
\end{algorithm}

\begin{algorithm}[htb]
\begin{algorithmic}
\STATE 
\newt{{\textbf{Given inputs}}:  $\vect{\afterchannel}$, $\epsilon$, $[\beta_1,...,\beta_\nturbo]$.}
\STATE 
{\textbf{Initialization}}: Set $p_D(\beforechannel_\iter)=\frac{1}{\modsize}\sum_{\beforechannel\in\mathcal{A}} \delta(\beforechannel_\iter-\beforechannel)$ for $\iter=1,\hdots, \Ntx$. \\
\notajj{\textbf{Turbo Iteration}: }
\FOR {$t=\noteISV{0},...,\nturbo$}
\STATE
\newt{
1) Compute the mean $\mu_{t_\iter}^{[1]}$ and variance $\sigma_{t_\iter}^{2[1]}$ of $p_D(\beforechannel_\iter)$. 
\STATE
{\textbf{Self Iteration}: }
\FOR {$\ell=1,...,\iterep$} 
\STATE
2) 
Run the moment matching procedure in \ALG{MMD} with inputs $p_D(\beforechannel_\iter)$, $\mu_{t_\iter}^{[\ell]},\sigma_{t_\iter}^{2[\ell]}$, $\vect{\afterchannel}$, $\epsilon$, $\beta_t$, to obtain  $\sigma_{t_\iter}^{2[\ell+1]}$ and $\mu_{t_\iter}^{[\ell+1]}$. 
\ENDFOR 
}
\\
\STATE
3) With the values $\sigma_{t_\iter}^{2[\iterep+1]}, \mu_{t_\iter}^{[\iterep+1]}$ obtained, calculate the extrinsic distribution $q_E^{[\iterep+1]}(\beforechannel_\iter)$ as in \EQ{extrinsicnubep}. 
\STATE
4) Demap the extrinsic distribution and compute the extrinsic \ac{LLR}, \newt{$L_E(b_n)$, \noteISV{and deliver it to the channel decoder}}. 
\STATE
5) Run the channel decoder to output \newt{$p_D(\beforechannel_\iter)$}. 
\ENDFOR
\end{algorithmic}
\caption{nuBEP Turbo Decoder for \ac{MIMO}}\LABALG{nuEPalgorithmMIMO}
\end{algorithm}

\subsection{EP parameters}

The update of the EP solution is a critical issue due to instabilities, particularly for high-order modulations. In this subsection, we review the EP parameters used in related approaches \cite{Senst11,Cespedes17} to propose some values. These parameters are: the minimum allowed variance ($\epsilon$), a damping factor ($\beta$) and the number of EP iterations ($\iterep$). The first two parameters determine the speed of the algorithm to get a stationary solution and control instabilities. The computational complexity of the algorithm depends linearly with $\iterep$. 

In \cite{Cespedes17}, the authors set $\iterep=10$ and introduced fast updates of EP solution by setting $\beta=0.95$. To avoid instabilities due to the fast updates, they set a gradual decrease for the minimum variance, starting with a high value along the first 4 iterations and then decreasing it exponentially \notajj{as} $\epsilon=2^{-\max(\ell-4,1)}$. However, we found that for large-size modulations and turbo schemes, the fast updates can provoque instabilities, as it will be showed in \SEC{sim}. For this reason and following our proposal in \cite{Santos18}, we let $\beta$ start with a conservative value and \notajj{increase} it exponentially with the number of turbo iterations, \newt{$\beta_{\noteISV{t}}$}$=\min(\exp({t/1.5})/10,0.7)$, where $t\in[0,\nturbo]$ is the number of the current turbo iteration. This growth of $\beta$ allows to reduce the number of EP iterations once the turbo procedure starts. We propose to reduce it from 10 in \cite{Cespedes17} to $\iterep=3$. We also set $\epsilon=10^{-8}$. Regarding the control of negative variances, we just update the EP solution when the computed variance is positive (see \EQ{negativevariance}), as proposed in \cite{Cespedes17}. 

\notajj{In \cite{Senst11} only one iteration of the EP procedure is computed, i.e.,  $\iterep=1$. They do not} introduce any damping or control of minimum variances. 
\notajj{In case of negative} variances, they update the EP solution with the moments computed in Step 3 of \noteISV{\ALG{MMD}}, i.e., 
\begin{align}
\sigma_{\noteISV{t}_{\iter}}^{2[\ell+1]}=\sigma_{p_\iter}^{2[\ell]}, \,\,\,\,\,\,\,\, \mu_{t_\iter}^{[\ell+1]}=\mu_{p_\iter}^{[\ell]}. 
\end{align}

In \TAB{parameters}, we describe the values of the EP parameters used in the current proposal (nuBEP) and the other EP-based detectors in the literature. \noteISV{The computational complexity per turbo iteration of these algorithms is included in \TAB{compl}.} \newt{The computational complexity of the MPEP \noteISV{\cite{Senst11}}, EPD \noteISV{\cite{Cespedes17}} and nuBEP is $S+1$ times the complexity of the LMMSE, of cubic order with $N_t$, where $S=1$ for the MPEP, $S=10$ for the EPD and $S=3$ for the nuBEP.}

\begin{table}[htb]
\begin{center}
\begin{tabular}{c c c c c}
\toprule
Algorithm & $\epsilon$ & $\beta$ & $\iterep$    \\
\midrule
nuBEP &  $1e^{-8}$ & $\min(\exp{(t/1.5)}/10,0.7)$ & 3  \\
EPD \cite{Cespedes17} & $2^{-\max(\ell-4,1)}$ & 0.95 & 10\\
\newt{MPEP} \cite{Senst11} & - & - & 1 \\
\bottomrule
\end{tabular}
\captionof{table}{\small Values for the EP parameters. }\LABTAB{parameters}
\end{center}
\end{table}

\begin{table}[htb]
\begin{center}
\begin{tabular}{c c }
\toprule
Algorithm & Computational Complexity Order  \\
\midrule
MPEP \noteISV{\cite{Senst11}}  & $2\order(N_t^3)$ \\
EPD \noteISV{\cite{Cespedes17}} &  $11\order(N_t^3)$ \\
nuBEP & $4\order(N_t^3)$\\
\bottomrule
\end{tabular}
\captionof{table}{\small Computational complexity order. }\LABTAB{compl}
\end{center}
\end{table}

\section{Simulation Results}\LABSEC{sim}

In this section we illustrate the performance of the proposed nuBEP turbo detector and compare its performance \notajj{to the ones of the EP-based detectors} proposed in \cite{Cespedes17} and \cite{Senst11}, \notajj{hereafter denoted as} EPD and MPEP, respectively. We also depict the BER of the LMMSE. We do not include the \ac{SD} \cite{Studer08}, \ac{MCMC} \cite{Datta13}, \ac{GTA} \cite{Goldberger11} or \ac{CHEMP} \cite{Narasimhan14} algorithms in the simulations because it has already been showed that the EPD \cite{Cespedes17} quite outperforms these three approaches \cite{Cespedes14,Cespedes17thesis}. 
The modulator uses a Gray mapping and a 128-QAM constellation. The results are averaged over $100$ random channels and $10^4$ random encoded words of length $\notaI{\t}=4096$ (per channel realization). A number of $\nturbo=5$ turbo iterations were run. Each channel \noteISV{coefficient} is independent and identically Gaussian distributed with zero mean and \noteISV{unit} variance. A (3,6)-regular \ac{LDPC} of rate $1/2$ is used. The absolute value of LLRs given to the decoder is limited to $5$ in order to avoid very confident probabilities. The decoder runs a maximum of $100$ iterations. 

In \FIG{128QAM6x6} we \notajj{include the BER obtained for a system} with $\Ntx=\Nrx=6$ antennas. It can be \notajj{observed that the EPD \cite{Cespedes17} improves the performance of the LMMSE but it is far from the results of the nuBEP and the} MPEP. The reason is that the true prior used \noteISV{by EPD} in the moment matching procedure is set to a uniform distribution, while nuBEP and MPEP use a non-uniform one given by the output of the decoder, which better \notajj{characterizes} the prior after the turbo feedback. The MPEP approach \cite{Senst11} quite outperforms both the LMMSE and the EPD because it uses a non-uniform prior. However, it does not achieve the performance of the nuBEP because it just computes one iteration of the EP algorithm and does not use any damping procedure. \notajj{The new proposed approach}, nuBEP, \notajj{exhibits} the \notajj{most} accurate and robust performance, due to its carefully chosen EP parameters, having gains of 14 dB and 6 dB with respect to the LMMSE and MPEP, respectively. 


\begin{figure}[t!]
\scalebox{0.9}{
%
%
%
\definecolor{mycolor1}{rgb}{0.00000,1.00000,1.00000}%

\definecolor{mycolor}{cmyk}{1,0,1,0}
\definecolor{mycolor2}{rgb}{0.00000,0.80000,1.00000}%
\definecolor{mycolor3}{rgb}{1.00000,0.00000,1.00000}%
\definecolor{mycolor4}{rgb}{0.45, 0.31, 0.59}%
\definecolor{mycolor5}{rgb}{0.6, 0.4, 0.8}
\definecolor{carnationpink}{rgb}{1.0, 0.65, 0.79}
\definecolor{auburn}{rgb}{0.43, 0.21, 0.1}
\begin{tikzpicture}[scale=1]


\begin{axis}[%
width=2.9in,
height=2in,
scale only axis,
every axis/.append style={font=\small},
xmajorgrids,
xmin=21,
xmax=45,
xlabel style={align=center}, xlabel={$\noteISV{\Ntx}\EsNo$ (dB)},
ymode=log,
ymin=1e-05,
ymax=2e-1,
yminorticks=true,
ylabel={BER},
name=plot6,
ymajorgrids,
yminorgrids,
legend style={at={(1,1)},anchor=north east,draw=black,fill=white,legend cell align=left,
,font=\footnotesize
}
]

\addplot [color=blue,solid,mark=triangle,mark size=2.2,line width=1.1,mark options={solid,,rotate=180}]
  table[row sep=crcr]{
20	0	\\
21	0	\\
22	0	\\
23	0	\\
24	0.0854520344842105	\\
25	0.0614420814884211	\\
26	0.0476901278105263	\\
27	0.0362831521473684	\\
28	0.0276537286526316	\\
29	0.0236935723473684	\\
30	0.0199732407052631	\\
31	0.0167442046947368	\\
32	0.0141069666105263	\\
33	0.0130041496631579	\\
34	0.0115634882105263	\\
35	0.00934657775789474	\\
36	0.00768540597894737	\\
37	0.00639932867368421	\\
38	0.00515281604210526	\\
39	0.00444998007368421	\\
40	0.00327315378947368	\\
41	0.00166798381052632	\\
42	0.000293883442105263	\\
43	1.66530526315789e-06	\\
44	0	\\
45	0	\\
46	0	\\
47	0	\\
};
\addlegendentry{LMMSE};

\addplot [color=red,solid,mark=square,mark size=2.0,line width=1.1,mark options={solid}]
  table[row sep=crcr]{
23	0.0905246925578947	\\
24	0.063217671031579	\\
25	0.0437931669578947	\\
26	0.0313147625263158	\\
27	0.0219820874736842	\\
28	0.0162878020105263	\\
29	0.012823893768421	\\
30	0.00988933470526316	\\
31	0.00725283313684211	\\
32	0.00577345004210526	\\
33	0.00472160858947368	\\
34	0.00365914728421053	\\
35	0.00259243667368421	\\
36	0.00137649126315789	\\
37	0.000597188294736842	\\
38	0.000162346326315789	\\
39	8.69342842105263e-05	\\
40	1.08860315789474e-05	\\
41	1.77942105263158e-05	\\
42	0	\\
43	6.22945263157895e-06	\\
44	0	\\
45	0	\\
46	0	\\
47	0	\\
};
\addlegendentry{EPD \cite{Cespedes17}};

\addplot [color=mycolor3,solid,mark=o,mark size=2.2,line width=1.1,mark options={solid}]
  table[row sep=crcr]{
20	0.120759392747368	\\
21	0.0781777442736842	\\
22	0.0408102266315789	\\
23	0.0232815510631579	\\
24	0.0130062578105263	\\
25	0.00645352375789474	\\
26	0.00257016670526316	\\
27	0.000796026147368421	\\
28	0.000206240178947368	\\
29	2.93632736842105e-05	\\
30	9.76557894736842e-08	\\
31	8.37768421052632e-06	\\
32	0	\\
33	0	\\
34	0	\\
35	0	\\
36	0	\\
37	0	\\
38	0	\\
39	0	\\
40	0	\\
41	0	\\
42	0	\\
43	0	\\
44	0	\\
45	0	\\
46	0	\\
47	0	\\
};
\addlegendentry{nuBEP};

\addplot [color=orange,solid,mark=+,mark size=2.5,line width=1.1,mark options={solid}]
  table[row sep=crcr]{
20	0.135671736574316	\\
21	0.0947544043894737	\\
22	0.0568462022315789	\\
23	0.0346656610105263	\\
24	0.0224093607484211	\\
25	0.0151532412210526	\\
26	0.00957894271578948	\\
27	0.00596657799789474	\\
28	0.00362434948421053	\\
29	0.00198992598947368	\\
30	0.000735154894736842	\\
31	0.000353104	\\
32	0.000159508568421053	\\
33	4.15503157894737e-05	\\
34	3.74075263157895e-05	\\
35	1.53631157894737e-05	\\
36	2.40547368421053e-06	\\
37	0	\\
38	0	\\
39	0	\\
40	0	\\
41	0	\\
42	0	\\
43	0	\\
44	0	\\
45	0	\\
46	0	\\
47	0	\\
};
\addlegendentry{MPEP \cite{Senst11}};

\end{axis}

\end{tikzpicture}%
}
\caption{\small BER along $\noteISV{\Ntx}\EsNo$ for nuBEP, EPD \cite{Cespedes17}, MPEP \cite{Senst11} and LMMSE detectors, 128-QAM and averaged over 100 randomly channels in a $6\times6$ system after \notaI{$\nturbo=5$ turbo} iterations.}
\LABFIG{128QAM6x6}
\end{figure}

In \FIG{128QAM32x32} we increase the number of antennas to $\Ntx=\Nrx=32$. In this scenario, the EPD approach shows instabilities at large $\noteISV{\Ntx}\EsNo$ since its parameters are not optimized for large-scale constellations and turbo schemes. Again, the best performance is obtained with the \notajj{nuBEP}, that has a remarkable improvement of 8 dB with respect to the LMMSE and of 1.5 dB compared to the MPEP algorithm. 

\begin{figure}[t!]
\scalebox{0.9}{
%
%
%
\definecolor{mycolor1}{rgb}{0.00000,1.00000,1.00000}%

\definecolor{mycolor}{cmyk}{1,0,1,0}
\definecolor{mycolor2}{rgb}{0.00000,0.80000,1.00000}%
\definecolor{mycolor3}{rgb}{1.00000,0.00000,1.00000}%
\definecolor{mycolor4}{rgb}{0.45, 0.31, 0.59}%
\definecolor{mycolor5}{rgb}{0.6, 0.4, 0.8}
\definecolor{carnationpink}{rgb}{1.0, 0.65, 0.79}
\definecolor{auburn}{rgb}{0.43, 0.21, 0.1}
\begin{tikzpicture}[scale=1]


\begin{axis}[%
width=2.9in,
height=2in,
scale only axis,
every axis/.append style={font=\small},
xmajorgrids,
xmin=20,
xmax=34,
xlabel style={align=center}, xlabel={$\noteISV{\Ntx}\EsNo$ (dB)},
ymode=log,
ymin=1e-05,
ymax=2e-1,
yminorticks=true,
ylabel={BER},
name=plot6,
ymajorgrids,
yminorgrids,
legend style={at={(1,1)},anchor=north east,draw=black,fill=white,legend cell align=left,
,font=\footnotesize
}
]

\addplot [color=blue,solid,mark=triangle,mark size=2.2,line width=1.1,mark options={solid,,rotate=180}]
  table[row sep=crcr]{
18	0	\\
19	0	\\
20	0	\\
21	0	\\
22	0	\\
25	0.13901182231579	\\
26	0.0994615178736842	\\
27	0.0498146594631579	\\
28	0.0163020739263158	\\
29	0.00330360183473684	\\
30	0.000383444631578947	\\
31	2.67472947368421e-05	\\
32	9.66284210526316e-07	\\
33	0	\\
34	0	\\
35	0	\\
36	0	\\
37	0	\\
38	0	\\
39	0	\\
40	0	\\
41	0	\\
};
\addlegendentry{LMMSE};

\addplot [color=red,solid,mark=square,mark size=2.0,line width=1.1,mark options={solid}]
  table[row sep=crcr]{
23	0.136418232105263	\\
24	0.105882686526316	\\
25	0.0569907551263158	\\
26	0.0151560031894737	\\
27	0.00202034307368421	\\
28	0.000213245505263158	\\
29	0.00030306787368421	\\
30	0.000679117094736842	\\
31	0.000244716557894737	\\
32	8.98442947368421e-05	\\
33	1.49207157894737e-05	\\
34	0	\\
35	0	\\
36	0	\\
37	0	\\
38	0	\\
39	0	\\
40	0	\\
41	0	\\
};
\addlegendentry{EPD \cite{Cespedes17}};

\addplot [color=mycolor3,solid,mark=o,mark size=2.2,line width=1.1,mark options={solid}]
  table[row sep=crcr]{
18	0.223540897894737	\\
19	0.200392487368421	\\
20	0.161381334526316	\\
21	0.0615762462736842	\\
22	0.00464711517894737	\\
23	7.84642842105263e-05	\\
24	0	\\
25	0	\\
26	0	\\
27	0	\\
28	0	\\
29	0	\\
30	0	\\
31	0	\\
32	0	\\
33	0	\\
34	0	\\
35	0	\\
36	0	\\
37	0	\\
38	0	\\
39	0	\\
40	0	\\
41	0	\\
};
\addlegendentry{nuBEP};

\addplot [color=orange,solid,mark=+,mark size=2.5,line width=1.1,mark options={solid}]
  table[row sep=crcr]{
21	0.134185700410526	\\
22	0.0416237723578947	\\
23	0.00444629424210526	\\
24	0.000241287715789474	\\
25	4.25063157894737e-06	\\
26	0	\\
27	0	\\
28	0	\\
29	0	\\
30	0	\\
31	0	\\
32	0	\\
33	0	\\
34	0	\\
35	0	\\
36	0	\\
37	0	\\
38	0	\\
39	0	\\
40	0	\\
41	0	\\
};
\addlegendentry{MPEP \cite{Senst11}};

\end{axis}

\end{tikzpicture}%
}
\caption{\small BER along $\noteISV{\Ntx}\EsNo$ for nuBEP, EPD \cite{Cespedes17}, MPEP \cite{Senst11} and LMMSE detectors, 128-QAM and averaged over 100 randomly channels in a $32\times32$ system after  \notaI{$\nturbo=5$ turbo} iterations.}
\LABFIG{128QAM32x32}
\end{figure}

\noteISV{Finally, in \FIG{128QAM32x32Noise} we include some results for the case of imperfect channel state information (CSI). Each coefficient of the channel matrix is i.i.d. generated as $\widehat{h}_{k,j}=h_{k,j}+\delta h_{k,j} $ where $\delta h_{k,j}\sim  \cgauss{\delta h_{k,j}}{0}{\sigma_H^2}$. We set $\sigma_H^2=10^{-3}$. 
In solid lines we represent the BER of the algorithms as described in the previous section, where the nuBEP exhibits a good and robust behavior except for high $\EsNo$. The results in dashed lines were obtained by modelling the effect of the error in the channel estimation as i.i.d. zero-mean Gaussian distributed noise of variance $\Ntx\sigma_H^2\energy$. Accordingly, we replaced 
$\sigma_\noise^2$ by $\Ntx\sigma_H^2\energy+\sigma_\noise^2$ in the algorithms. It can be observed that the BER quite improves, avoiding the degradation in the performance at large $\EsNo$. Also, the result in dashed line for the nuBEP is close to the one in \FIG{128QAM32x32}, i.e., when no error is introduced in the channel matrix. More complex covariances for the noise could be introduced \cite{Ghavami17}.
}


\begin{figure}[t!]
\scalebox{0.9}{
%
%
%
\definecolor{mycolor1}{rgb}{0.00000,1.00000,1.00000}%

\definecolor{mycolor}{cmyk}{1,0,1,0}
\definecolor{mycolor2}{rgb}{0.00000,0.80000,1.00000}%
\definecolor{mycolor3}{rgb}{1.00000,0.00000,1.00000}%
\definecolor{mycolor4}{rgb}{0.45, 0.31, 0.59}%
\definecolor{mycolor5}{rgb}{0.6, 0.4, 0.8}
\definecolor{carnationpink}{rgb}{1.0, 0.65, 0.79}
\definecolor{auburn}{rgb}{0.43, 0.21, 0.1}
\begin{tikzpicture}[scale=1]


\begin{axis}[%
width=2.9in,
height=2in,
scale only axis,
every axis/.append style={font=\small},
xmajorgrids,
xmin=20,
xmax=29,
xlabel style={align=center}, xlabel={$\Ntx\EsNo$ (dB)},
ymode=log,
ymin=1e-05,
ymax=2e-1,
yminorticks=true,
ylabel={BER},
name=plot6,
ymajorgrids,
yminorgrids,
legend style={at={(0,0)},anchor=south west,draw=black,fill=white,legend cell align=left,
,font=\footnotesize
}
]

\addplot [color=blue,solid,mark=triangle,mark size=2.2,line width=1.1,mark options={solid,,rotate=180}]
  table[row sep=crcr]{
20	0	\\
21	0	\\
22	0	\\
25	0.1719979	\\
26	0.156519840631579	\\
27	0.136216217852632	\\
28	0.112456318126316	\\
29	0.0895513248842105	\\
30	0.0726450288210526	\\
31	0.0648434420905263	\\
32	0.0638056125789474	\\
33	0.0706702712736842	\\
};
\addlegendentry{LMMSE};

\addplot [color=red,solid,mark=square,mark size=2.0,line width=1.1,mark options={solid}]
  table[row sep=crcr]{
23	0.155284452631579	\\
24	0.137167736	\\
25	0.114976873157895	\\
26	0.0862009899157895	\\
27	0.0609750964	\\
28	0.0695999907789474	\\
29	0.116152310421053	\\
30	0.138439884421053	\\
31	0.140774302842105	\\
32	0.136065358842105	\\
33	0.130813432936842	\\
};
\addlegendentry{EPD \cite{Cespedes17}};

\addplot [color=mycolor3,solid,mark=o,mark size=2.2,line width=1.1,mark options={solid}]
  table[row sep=crcr]{
20	0.182699082421053	\\
21	0.123722735978947	\\
22	0.0381712781326316	\\
23	0.00526793187368421	\\
24	0.000651879789473684	\\
25	7.39825473684211e-05	\\
26	2.02505263157895e-06	\\
27	0	\\
28	0	\\
29	0	\\
};
\addlegendentry{nuBEP};

\addplot [color=orange,solid,mark=+,mark size=2.5,line width=1.1,mark options={solid}]
  table[row sep=crcr]{
21	0.170187582736842	\\
22	0.112060276284211	\\
23	0.0470066863157895	\\
24	0.0162807647263158	\\
25	0.006411074	\\
26	0.00284405264210526	\\
27	0.00176807833684211	\\
28	0.00172750502105263	\\
29	0.00187897033684211	\\
30	0.00276679888421053	\\
31	0.00528062374736842	\\
32	0.00945819995789473	\\
33	0.0158676190936842	\\
};
\addlegendentry{MPEP \cite{Senst11}};

\addplot [color=mycolor3,solid,mark=o,mark size=2.2,line width=1.1,mark options={solid}]
  table[row sep=crcr]{
30	1.06494736842105e-05	\\
31	6.84008315789474e-05	\\
32	0.000689835810526316	\\
33	0.00436357394947368	\\
};

\addplot [color=blue,dashed,mark=triangle,mark size=2.2,line width=1.1,mark options={solid,,rotate=180}]
  table[row sep=crcr]{
20	0	\\
21	0	\\
22	0	\\
25	0.167970078947368	\\
26	0.152208235789474	\\
27	0.131188932473684	\\
28	0.106006094210526	\\
29	0.0785349250526316	\\
30	0.0516333428421053	\\
31	0.0342113968421053	\\
32	0.0210041472105263	\\
33	0.0128001491578947	\\
};

\addplot [color=red,dashed,mark=square,mark size=2.0,line width=1.1,mark options={solid}]
  table[row sep=crcr]{
23	0.154694373684211	\\
24	0.136350659473684	\\
25	0.113752589473684	\\
26	0.0828562660526316	\\
27	0.0473076959242105	\\
28	0.020297641	\\
29	0.00757491468421053	\\
30	0.00193893821052632	\\
31	0.000823141736842105	\\
32	0.000288804368421053	\\
33	0.000415704684210526	\\
};

\addplot [color=mycolor3,dashed,mark=o,mark size=2.2,line width=1.1,mark options={solid}]
  table[row sep=crcr]{
20	0.182043515789474	\\
21	0.121953670631579	\\
22	0.0309534258421053	\\
23	0.00234626963157895	\\
24	0.000158819315789474	\\
25	0	\\
26	0	\\
27	0	\\
28	0	\\
29	0	\\
30	0	\\
31	0	\\
32	0	\\
33	0	\\
};

\addplot [color=orange,dashed,mark=+,mark size=2.5,line width=1.1,mark options={solid}]
  table[row sep=crcr]{
21	0.167778989473684	\\
22	0.102381078263158	\\
23	0.0295031668421053	\\
24	0.00494785026315789	\\
25	0.000532894684210526	\\
26	6.43763157894737e-05	\\
27	0	\\
28	0	\\
29	0	\\
30	0	\\
31	0	\\
32	0	\\
33	0	\\
};

\end{axis}

\end{tikzpicture}%
}
\caption{\small \noteISV{BER along $\Ntx\EsNo$ for nuBEP, EPD \cite{Cespedes17}, MPEP \cite{Senst11} and LMMSE detectors, 128-QAM and averaged over 100 randomly noisy channels with $\sigma_H^2=10^{-3}$ in a $32\times32$ system after $\nturbo=5$ turbo iterations. Solid lines correspond to a noise variance of $\sigma_\noise^2$, while dashed lines to $\Ntx\sigma_H^2\energy+\sigma_\noise^2$. } }
\LABFIG{128QAM32x32Noise}
\end{figure}
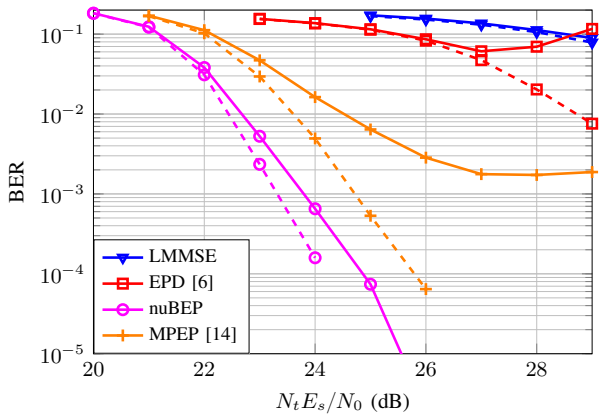

\section{Conclusions}\LABSEC{conc}

We have proposed an EP-based turbo detector (nuBEP) for MIMO systems and large-size modulations where the optimal MAP algorithm is computationally unfeasible. The nuBEP detector quite outperforms the classical LMMSE and other EP-based detectors found in the literature. Specifically, it uses a non-uniform prior, rather than a uniform one as in \cite{Cespedes17}. This prior better characterizes the true prior used in the self-iterations of the EP algorithm, during the moment matching procedure, once the turbo procedure has started. \notaI{The proposed detector also optimizes its parameters to avoid some instabilities that appear at large $\EsNo$ and to reduce its complexity. Specifically, it reduces the number of EP iterations from 10 (used in  \cite{Cespedes17}) to 3 after the feedback from the decoder.} It also outperforms the EP detector in \cite{Senst11} since we include a self-iterated stage with damping and a different control of negative variances, that endow the nuBEP approach with a more accurate solution. 
Simulations results show that the proposed nuBEP turbo detector has gains \notajj{in the range} 5-11 dB with respect to the EPD \cite{Cespedes17} and 1.5-6 dB compared to the MPEP \cite{Senst11}.



%

%


%
%

\ifCLASSOPTIONcaptionsoff
  \newpage
\fi



%
%
%

\bibliographystyle{IEEEtran}
\bibliography{allBib}

%

%
%
%




\end{document}